\documentstyle[12pt]{article}

\newcommand{\be}{\begin{equation}}
\newcommand{\ee}{\end{equation}}
\newcommand{\sB}{\stackrel{\rightarrow}{B}}
\newcommand{\sS}{\stackrel{\rightarrow}{S}}
\newcommand{\sn}{\stackrel{\rightarrow}{n}}
\newcommand{\sr}{\stackrel{\rightarrow}{r}}
\newcommand{\sse}{\stackrel{\rightarrow}{e}}  
\newcommand{\sH}{\stackrel{\rightarrow}{H}}  
\newcommand{\om}{\omega}  
\newcommand{\gm}{\gamma} 
\newcommand{\ra}{\rightarrow}

\begin{document}

\begin{center}
{\Large{\bf Spin Maser under Stationary Pumping} \\ [5mm]
V. I. Yukalov and E. P. Yukalova} \\ [3mm]

{\it Centre for Interdisciplinary Studies in Chemical Physics \\
University of Western Ontario, London, Ontario N6A 3K7, Canada \\
and \\
Bogolubov Laboratory of Theoretical Physics \\
Joint Institute for Nuclear Research, Dubna 141980, Russia}
\end{center}

\vspace{3cm}

\begin{abstract}

Spin dynamics of a polarized spin system is studied when the latter is 
coupled with a resonant electric circuit and is under the action of an 
external pumping supporting a stationary nonequilibrium magnetization. A 
complete classification of possible regimes of spin motion is given. In 
addition to seven regimes considered earlier, two other transient regimes 
are found and thoroughly described: One is an oscillatory regime, when spins 
always move coherently but the degree of coherence fluctuates with time.
Another is a pulsing regime, when spins reveal coherent motion during short 
pulses separated from each other by intervals of incoherent motion. These
regimes are, in principle, transient, although may be extremely long lasting;
their duration may be several orders longer than the transverse relaxation 
time and twice longer than the longitudinal relaxation time. Both transient 
regimes end with a coherent quasistationary regime.

\end{abstract}

\newpage

\section{Introduction}

Nonequilibrium resonance phenomena in spin systems have their counterparts in 
atomic systems. Recall, for instance, free induction, occurring similarly in 
both kinds of the systems, or spin echo being a direct analog of photon echo.
The feasibility of a self--organized coherent process, called superradiance, 
has been theoretically predicted almost simultaneously for spin [1] and atomic
[2] systems. The difference between these is that for realizing such a coherent
process in a spin system, the latter must be coupled with a resonator. 
Superradiance in atomic and molecular systems has been studied, both
theoretically and experimentally, quite in detail, and has been expounded in a 
number of reviews and books, of which we cite only some recent Refs. [3-6].

In analogy with atomic superradiance, the process of collective coherent 
relaxation in spin systems has been called spin superradiance. This process
was observed for electron [7-9] as well as for nuclear spins [10-12]. 
Accurate experiments observing purely self--organized superradiance from 
proton spins have been accomplished [13-16]. The peculiarities of spin 
superradiance were studied by means of computer simulation [17,18], being 
based on the microscopic Hamiltonian of a nuclear magnet, commonly accepted 
in the theory of nuclear magnetic resonance [19]. An analytic theory for this
microscopic model was developed in Refs. [20,21], where it was shown that, 
despite many similarities between spin and optic superradiance, there are
also crucial differences between them. For instance, because of the principal
role in triggering the radiation process, played by direct dipole 
interactions, the self--organized coherent relaxation in nuclear magnets 
is of non--Dicke type [21,22]. The behaviour of resonant spin systems under 
the action of an injected signal with particularly chosen delay times [23] 
and under the influence of parametric excitations [24-26] has also been 
considered. The importance of studying nonequilibrium coherent phenomena in
spin systems is caused by the usage of these phenomena for many applications,
for example, for spin masers [27-30], for the repolarization of scattering
targets [16,31], and for a possible creation of sensitive particle detectors 
[32].

There is a problem that has not yet been properly studied for nuclear spin
systems coupled with a resonator: What would be the behaviour of such a 
system under the influence of a stationary nonresonant pumping supporting a
nonequilibrium magnetization? This kind of pumping could be realized by means 
of dynamical nuclear polarization. Note that an equivalent question was posed
for electron spin systems [33] and considered in the framework of the 
adiabatic approximation. However, the latter, as is well known [34], is valid 
only at the last stage of relaxation, when the system is already close to its
stationary state. The adiabatic approximation cannot describe a transient
process, as has been discussed in detail in Refs. [21,35]. Therefore, the 
authors of Ref. [33] considered only the asymptotic stationary regime for a
spin system with a constant nonresonant pumping. Such a problem for atomic
systems has been analysed earlier and it has been shown that in the presence
of a constant external pumping atomic systems may exhibit only pulsed 
operation and cannot work in the quasistationary regime (see discussion in 
Refs. [36,37]).

In the present paper, we consider a nuclear magnet coupled with a resonant 
electric circuit and subject to the action of a constant nonresonant pumping 
supporting a nonequilibrium stationary magnetization. The consideration is 
based on the standard microscopic Hamiltonian [19] with dipole interactions 
between nuclear spins. We do not invoke the adiabatic approximation, but use 
the scale separation approach [20,21]. Therefore, we may analyse all possible 
transient regimes.

\section{Nuclear Magnet}

The standard Hamiltonian modelling a solid sample consisting of $N$ nuclear 
spins can be written [19] in the form
\be
\hat H =\frac{1}{2}\sum_{i\neq j}^N H_{ij} -\mu\sum_{i=1}^N\sB\cdot\sS_i ,
\ee
in which
$$ H_{ij}=\frac{\mu^2}{r_{ij}^3}\left [\sS_i\cdot\sS_j -3\left (\sS_i\cdot
\sn_{ij}\right)\left (\sS_j\cdot\sn_{ij}\right )\right ] $$
is the dipole interaction energy; $\mu$, a nuclear magneton; $\sS_i=\{
S_i^x,S_i^y,S_i^z\}$, a spin operator; and
$$ r_{ij}\equiv |\sr_{ij}|, \qquad \sn_{ij}\equiv\frac{\sr_{ij}}{r_{ij}} ,
\qquad \sr_{ij}\equiv \sr_i -\sr_j . $$
The total magnetic field
\be
\sB =\sH_0 +\sH, \qquad \sH_0 =H_0\sse_z , \qquad \sH= H\sse_x ,
\ee
consists of an external magnetic field $\sH_0$ directed along the $z$ axis 
and of a feedback field $\sH$ of the coil of a resonator electric
circuit with the coil axis being directed along the axis 
$x$. The resonator magnetic field $H$ is formed by an electric current 
satisfying the Kirchhoff equation.

Let us define the Larmor frequency $\om_0$ and the resonator frequency $\om$, 
being, respectively,
\be
\om_0\equiv\frac{\mu H_0}{\hbar} , \qquad \om\equiv\frac{1}{\sqrt{LC}} ,
\ee
where $H_0$ is assumed to be positive, $L$ is the coil inductance, and $C$ is the circuit capacity. Introduce also the resonator ringing damping.
\be
\gm_3\equiv\frac{\om}{2Q} , \qquad Q\equiv \frac{\om L}{R} ,
\ee
in which $Q$ is the quality factor of a circuit and $R$, its resistance.
Then the Kirchhoff equation can be written in the form
\be
\frac{dH}{dt} +2\gm_3 H +\om^2\int_0^t H(t')dt' = -4\pi\eta\rho
\frac{dM_x}{dt} , 
\ee
where $\eta$ is the coil filling factor; $\rho$, the density of spins; and
$$ M_x =\frac{\mu}{N}\sum_{i=1}^N\langle S_i^x\rangle $$
is the $x$ component of the reduced magnetization.

There are the following characteristic times: the spin--lattice relaxation 
time $T_1$; the spin--spin dephasing time $T_2$; the inhomogeneous 
broadening time $T_2^*$, due to local random fluctuations; and the resonator 
ringing time $T_3$. The width corresponding to these times are
$$ \gm_1\equiv\frac{1}{T_1} , \qquad \gm_2\equiv\frac{1}{T_2} , \qquad
\gm_2^*\equiv\frac{1}{T_2^*} , \qquad \gm_3\equiv\frac{1}{T_3} . $$
In the presence of pumping with the pumping velocity $\gm_p$, the effective
longitudinal relaxation becomes
\be
\gm_1^* \equiv \gm_1 +\gm_p .
\ee
All these widths are usually small, as compared to the frequencies in (3), 
defining the set of small parameters
\be
\frac{\gm_1^*}{\om_0}\ll 1 , \qquad \frac{\gm_2}{\om_0}\ll 1 , \qquad
\frac{\gm_2^*}{\om_0}\ll 1 , \qquad \frac{\gm_3}{\om}\ll 1 .
\ee
The resonator natural frequency is tuned to be close to the Larmor frequency, 
so that the detuning from the resonance is small,
\be
\frac{|\Delta|}{\om_0}\ll 1 , \qquad \Delta \equiv \om -\om_0 .
\ee

The existence of small parameters in (7) and (8) justifies the use of the 
scale separation approach, whose all details have been thoroughly expounded 
in Ref. [21]. Employing this approach, we may derive the evolution equations 
for the transverse magnetization 
\be
u \equiv\frac{1}{N}\sum_{j=1}^N\langle S_j^x - iS_j^y\rangle
\ee
and the longitudinal magnetization
\be
z\equiv\frac{1}{N}\sum_{j=1}^N\langle S_j^z\rangle .
\ee
The coupling between the spin sample and the resonant electric circuit is 
described [20,21] by the effective coupling parameter
\be
g \equiv \pi^2\eta\left (\frac{\rho\mu^2}{\hbar\gm_2}\right )
\ee

It is convenient to introduce the function
\be
w \equiv v^2 - 2\left (\frac{\gm_2^*}{\om_0}\right )^2 z^2 , \qquad  
v\equiv |u| .
\ee
After averaging over the time $2\pi/\om_0$ of fast oscillations and over 
random local fields, we obtain [21] the system of equations for function (12),
\be
\frac{dw}{dt} = -2\gm_2\left ( 1 +gz\right ) w ,
\ee and for the longitudinal magnetization (10),
\be
\frac{dz}{dt} =g\gm_2 w -\gm_1^*\left ( z-\zeta\right ) .
\ee
The derivation of eqs. (13) and (14) has been carefully explained in 
Ref. [21]. The only difference, in the case we consider now, is that the 
spin--lattice relaxation constant $\gm_1$ is replaced by the effective 
longitudinal width $\gm_1^*=\gm_1 +\gm_p$, including the pumping velocity 
$\gm_p$, and that the stationary magnetization parameter $\zeta$, in the 
presence of pumping, becomes negative, $\zeta<0$. The evolution equations 
are complemented by the initial conditions
\be
w(0)=w_0, \qquad z(0)=z_0 .
\ee
In what follows we assume that the coupling parameter (11) is nonzero, 
since the case $g\ra 0$ would result in the trivial exponential relaxation 
of solutions to Eqs. (13) and (14).

The spin--lattice relaxation parameter $\gm_1$ is usually much less than the 
dephasing width $\gm_2$, although for some materials they can be rather close 
to each other. For instance, in the case of $^3He$ at low temperature 
$T\sim 1K$, for the characteristic times one has [38] the values $T_1= 100 -
300 s$ and $T_2=30-100s$, so that $\gm_1/\gm_2\sim 1/3$. And as far as the 
effective relaxation parameter in the presence of pumping is the sum 
$\gm_1^*=\gm_1 +\gm_p$, hence $\gm_1^*>\gm_1$, as a result of which 
$\gm_1^*$ can become comparable with $\gm_2$ even if $\gm_1\ll\gm_2$. 

What is more significant is that, even, when $\gm_1^*$ is negligibly small as
compared to $\gm_2$, the term in (14) containing $\gm_1^*$, as will be shown 
in what follows, cannot be omitted if $\zeta<0$, that is, if a pumping is
present. This situation is drastically different from the case when pumping 
is absent [20-22]. In the latter case, if $\gm_1^*\ll\gm_2$, then in the time 
interval $0\leq t\leq (\gm_1^*)^{-1}$ one can omit the last term in (13). 
This omition allows to solve the system of Eqs. (13) and (14) exactly, 
yielding
\be
w=\left (\frac{\gm_0}{g\gm_2}\right )^2{\rm sech}^2
\left (\frac{t-t_0}{\tau_0} \right )
\ee
and
\be
z =\frac{\gm_0}{g\gm_2}\tanh\left (\frac{t-t_0}{\tau_0}\right ) -
\frac{1}{g} , 
\ee
where the radiation width is
\be
\gm_0 =\gm_2\sqrt{(1+gz_0)^2 +g^2w_0} ,
\ee
the radiation time is $\tau_0=\gm_0^{-1}$, and the delay time is
\be
t_0 =\frac{\tau_0}{2}\ln\left |
\frac{\gm_0 -\gm_2(1+gz_0)}{\gm_0 +\gm_2(1+gz_0)}\right | .
\ee
Solutions (16) and (17), depending on initial conditions and the coupling 
parameter (11), describe seven qualitatively different regimes of spin 
relaxations: free induction, collective induction, free relaxation, 
collective relaxation, weak superradiance, pure superradiance, and triggered 
superradiance. This classification is valid when the pumping is absent, for 
which case all these regimes have been analyzed earlier [20-22].

In the presence of pumping, we may expect that solutions (16) and (17) 
correctly describe the beginning of the relaxation process for time 
$t\ll (\gm_1^*)^{-1}$. These solutions, when $0\leq\gm_1^*t_0\leq 1$, depict 
the first superradiant burst occurring at the time $t=t_0$, where
\be
w(t_0) =z^2(t_0) = w_0 +\left ( z_0 +\frac{1}{g}\right )^2 .
\ee
However, these solutions do not give the overall picture for all times, even
when $\gm_1^*\ll\gm_2$. As the analysis of the following sections shows, the 
whole behaviour of solutions to Eqs. (12) and (13), in the presence of 
pumping, is essentially more complicated.

\section{Stability Analysis}

In order to understand, from the mathematical point of view, why the 
solutions to Eqs. (13) and (14), if one puts $\gm_1^*$ zero, can be 
drastically different from those when $\gm_1^*\neq 0$, even if $\gm_1^*$ is 
negligibly small as compared to $\gm_2$, one has to accomplish the stability 
analysis. For this purpose, we write Eqs. (13) and (14) in the form
\be
\frac{dw}{dt} = V_1 , \qquad \frac{dz}{dt} = V_2 ,
\ee
in which
$$ V_1=-2\gm_2(1 +gz)w , \qquad V_2 =g\gm_2w -\gm_1^*(z-\zeta) . $$
The equations $V_1=V_2=0$ define stationary, or fixed, points. We have two 
such points, one is
\be
z_1^* =\zeta , \qquad w_1^* = 0
\ee
and another is
\be
z_2^* = -\frac{1}{g} , \qquad w_2^*=-\frac{\gm_1^*}{g^2\gm_2}\left ( 1 +
g\zeta \right ) .
\ee

The Jacobian matrix
\begin{eqnarray}
\hat J=\left [ \begin{array}{cc}
\frac{\partial V_1}{\partial w} & \frac{\partial V_1}{\partial z} \\   
\\
\frac{\partial V_2}{\partial w} & \frac{\partial V_2}{\partial z}  
\end{array} \right ] \; ,
\end{eqnarray}
corresponding to (21), takes the form
\begin{eqnarray}
\hat J=\left [ \begin{array}{cc}
-2\gm_2(1+gz) & -2\gm_2gw \\   
\\
g\gm_2 & -\gm_1^*  
\end{array} \right ] \; .
\end{eqnarray}
The eigenvalues of matrix (25) are
\be
\lambda^\pm =-\frac{1}{2}\left\{ \gm_1^* +2\gm_2(1 +gz ) \pm
\left [ \left ( \gm_1^* -2\gm_2 (1 +gz )\right )^2 - 
8g^2\gm_2^2w\right ]^{1/2}\right\} .
\ee
The values of (26) evaluated at the fixed points define the Lyapunov
exponents. At the first fixed point, given by (22), we have
\be
\lambda_1^+ =-\gm_1^* , \qquad \lambda_1^- = - 2\gm_2 (1 +g\zeta ) .
\ee
And at the second fixed point, given by (23), we find
\be
\lambda_2^\pm = -\frac{1}{2}\left\{ \gm_1^* \pm
\sqrt{(\gm_1^*)^2 +8\gm_1^*\gm_2(1 +g\zeta)}\right \} .
\ee

The stability of the fixed points and, consequently, the stability of motion 
is characterized by the signs of the Lyapunov exponents [39]. Varying the 
value of the pumping parameter $\zeta$, we may get qualitatively different 
regimes of motion. These regimes are separated by the pumping thresholds
\be
\zeta_1 \equiv -\frac{1}{g} , \qquad \zeta_2\equiv -\frac{1}{g}\left (
1 +\frac{\gm_1^*}{8\gm_2}\right ) .
\ee

When the pumping parameter satisfies the inequality
\be
\zeta > \zeta_1 ,
\ee
then
$$ \lambda_1^\pm < 0 , \qquad \lambda_2^+ < 0 , \qquad \lambda_2^- > 0 . $$
Hence, the fixed point (22) is a stable node, while that (23) is a saddle
point.

In the case when
\be
\zeta =\zeta_1 ,
\ee
we have
$$ \lambda_1^+ =\lambda_2^+ < 0 , \qquad \lambda_1^- =\lambda_2^- = 0 . $$
Both fixed points (22) and (23) merge together becoming neutrally stable.
The pumping threshold $\zeta_1$ corresponds to a bifurcation point.

When the pumping parameter is in the region                                                                   
\be
\zeta_2 \leq \zeta < \zeta_1 ,
\ee
then
$$ \lambda_1^+ < 0 , \qquad \lambda_1^- > 0 , \qquad \lambda_2^\pm < 0 , $$
which means that the fixed points interchange their properties: now (22) is 
a saddle point and (23) becomes a stable node.

For sufficiently strong pumping, when
\be
\zeta < \zeta_2 ,
\ee
the fixed point (22), as earlier, continues to be a saddle point, since 
$\lambda_1^+ < 0$ and $\lambda_1^- >0$. But for the fixed point (23) we get
\be
\lambda_2^\pm = -\frac{1}{2}\gm_1^* \pm i\Omega ,
\ee
with
\be
\Omega =\frac{1}{2}\sqrt{|(\gm_1^*)^2 +8\gm_1^*\gm_2(1+g\zeta)|} .
\ee
This shows that (23) transforms to a stable focus.

Let us notice that under all conditions, if one puts $\gamma_1^*\ra 0$, then
(27) and (28) yield $\lambda_1^+=0$ and $\lambda_2^\pm=0$. This tells us that
both fixed points correspond to a structurally unstable system [39]. 
Structural instability means that the temporal behaviour of the system can be 
essentially disturbed under an arbitrary small change of the evolution 
equations. This is why the term of Eq. (14) containing $\gm_1^*$, in general,
must not be omitted even if $\gm_1^*$ is many orders smaller than $\gm_2$. It
may happen that for time $t\ll (\gm_1^*)^{-1}$, one can omit this term in some
particular cases. For instance, this is the case of a spin maser without
pumping [20,21], that is, with $\zeta\geq 0$. However, for a spin maser in
the presence of pumping, when $\zeta <0$, the situation can be drastically 
different. In the latter case, to describe the temporal behaviour of the
system, one has to keep the term with $\gm_1^*$.

\section{Numerical Solution}

To analyse the behaviour of solutions to the system of equations (13) and 
(14), we solved this system numerically. The spin--resonator coupling 
parameter (11) is taken to be $g=10$. For a weak pumping, when 
$\zeta\geq\zeta_1=-0.1$, as well as for an intermediate pumping, when $\zeta$ 
satisfies (32), the behaviour of solutions is similar to that studied in Refs.
[20-22,28-30], which is caused by the fact that the stationary point in all
these cases is a stable node. The most interesting here is the case when
the fixed point is a stable focus. Then qualitatively new types of solutions 
appear. Therefore, in what follows we concentrate our attention on the case
of strong pumping corresponding to inequality (33). To this end, we take the
pumping parameter $\zeta=-0.5$. A few typical phase portraits for a system
with a fixed point being a stable focus are presented in Fig.1. For 
convenience, we introduce a notation
$$ \gm\equiv \frac{\gm_1^*}{\gm_2} . $$
Increasing the pumping velocity leads to the increase of $\gm$, as a result
of which the stationary value of $w_2^*$ also increases. Note that the 
initial value of $z_0=z(0)$ does not influence much the whole picture. The
phase portraits for $z_0=-0.5$ and $z_0=+0.5$ are very similar to each other.

The following figures show the temporal behaviour of the functions $w(t)$
and $z(t)$ for various initial conditions and pumping velocities. When there
are no external pumping fields, except that realizing a stationary dynamical
polarization $\zeta$, then we should put $z(0)=\zeta$. However, the initial
polarization $z_0=z(0)$ can be made different from $\zeta$ by using additional 
short external pulses. Keeping this possibility in mind, we consider the cases
with $z(0)$ not always coinciding with $\zeta$.

In Figs. 2 to 6, we see the oscillatory regime of motion. Everywhere, if it 
is not stated otherwise, we take $g=10$ and $\zeta=-0.5$. It is only in 
Fig. 7 where the pumping parameter is varyed. Fig. 8 demonstrates how an 
oscillatory regime of motion changes to a pulsing one with changing initial 
conditions. The pulsing regime of motion is presented in Figs. 9 to 12.

\section{Discussion}

We have considered the dynamics of spin maser under a stationary pumping 
supporting a constant nonequilibrium magnetization in a system of nuclear 
spins. Such a pumping can be accomplished by means of stationary dynamical 
polarization of nuclei. The regimes of oscillatory and pulsing motion are 
found. The distinction between these two regimes is, of course, somewhat 
conditional, depending on the level of coherence existing during the time
intervals separating the neighbouring coherent bursts. These bursts can be 
directly observed by measuring the current power $P$ that is proportional 
to $v^2$. As far as $\gm_2^*\ll\om_0$, we have $v^2\sim w$. Whence, $P\sim w$. 
In this way, the function $w(t)$ is proportional to the current power and, 
thus, is an observable quantity. This function is also connected with the
intensity of magnetodipole radiation, $I(t)$, and the coherence coefficient 
$C_{coh}(t)$, as is discussed in Refs. [17,18]. Therefore, the coherent
bursts occurring in the system of nuclear spins can be named superradiant.

Superradiant regimes appearing in a spin maser under the action of the
nonresonant pumping, supporting a stationary level of a nonequilibrium
magnetization $\zeta$, are quite different form the regimes developing under 
the influence of a resonant pumping realized by means of an alternating
external fields. For comparison, we adduce in Figs. 13 to 15 the behaviour
of the coherence coefficient $C_{coh}$, radiation intensity $I$, and of the
polarization $p_z\equiv -z(t)$ for a nuclear magnet pumped with a resonant 
alternating field [17,18] oscillating as $h_0\cos\om t$.

Dynamics of a spin maser in the presence of a nonresonant pumping supporting 
a pumping polarization $\zeta <\zeta_2$ resembles that of pulsing lasers 
[36,37]. The difference is that pulsing lasers cannot operate in a stationary 
regime, while a pumped spin maser, after its oscillatory or pulsing stage, 
tends to a stationary regime with a current power proportional to $w_2^*$ 
in (23). The level of coherence in this stationary regime, as compared to 
that of the first superradiant burst, defined in (20), is described by the 
ratio
$$ \frac{w_2^*}{w(t_0)} =
\frac{\gm_1^*|1+g\zeta|}{\gm_2(1+gz_0)^2 +g^2w_0} . $$
The latter, for $w_0=0, \; z_0\sim\zeta\sim 1$, and $g\gg 1$, gives
$$ \frac{w_2^*}{w(t_0)}\sim \frac{\gm_1^*}{\gm_2 g} . $$
Therefore, to reach the intensity of the first superradiant burst, the 
pumping velocity is to be very high, so that $\gm_1^* \sim \gm_2 g$ which
looks as practically unattainable.

It is worth paying some attention to terminology. The temporal behaviour of 
solutions, as is seen from above figures, is not periodic, since the time 
intervals between pulses as well as their amplitudes change with time. This
time dependence cannot be called quasiperiodic (in mathematical sense). It is
not chaotic too, although for some parameters and in a limited time interval 
it may look as pseudochaotic, slightly reminding quantum pseudochaos [40].
Therefore, the most appropriate adjectives characterizing the type of 
solutions we found could be, probably, oscillating and pulsing. Or we could 
describe all of them by one word, e.g., pulsing, implying that this
incapsulates all admissible variants of solutions consisting of a number of 
pulses. Such a pulsing operation can be employed in spin masers.

\vspace{5mm}

{\bf Acknowledgement}

\vspace{2mm}

We are very grateful to C. M. Bowden for useful discussions and comments.
Financial support from the University of Western Ontario is appreciated.

\newpage

\begin{center}
{\bf Figure Captions}
\end{center}

{\bf Fig.1.} The phase portraits for $g=10,\;\zeta=-0.5,\;w_0=0,001$, and for
the time interval $0\leq t\leq 100\gm_2^{-1}$ for different pumping 
velocities and initial polarizations: (a) $\gm=0.1,\; z_0=0.5$; (b) 
$\gm=1,\; z_0=-0.5$; (c) $\gm=1,\; z_0=0.5$.

\vspace{5mm}

{\bf Fig.2.} The temporal behaviour of solutions for $\gm=0.001$, with 
initial conditions $w_0=10^{-6}$ and $z_0=-0.1$: (a) $w(t)$; (b) $z(t)$.

\vspace{5mm}

{\bf Fig.3.} Solutions to evolution equations for $\gm=0.001$, with initial
conditions $w_0=10^{-6}$ and $z_0=-0.1$: (a) $w(t)$; (b) $z(t)$.

\vspace{5mm}

{\bf Fig.4.} Solutions of equations as functions of time for $\gm=1$ and 
initial conditions $w_0=0.001$ and $z_0=-0.5$: (a) $w(t)$; (b) $z(t)$.

\vspace{5mm}

{\bf Fig.5.} Spin dynamics for initial conditions $w_0=0.1$ and $z_0=-0.25$,
with different parameters $\gm$, where the solid line is for $\gm=1$ while 
the dashed line is for $\gm=0.5$: (a) $w(t)$; (b) $z(t)$.

\vspace{5mm}

{\bf Fig.6.} Spin dymanics with initial conditions $w_0=0.5,\; z_0=0.5$ for
$\gm=1$ (solid line) and $\gm=0.5$ (dashed line): (a) $w(t)$; (b) $z(t)$.

\vspace{5mm}

{\bf Fig.7.} Spin dynamics for $\gm=1$ with initial conditions $w_0=0.5,\;
z_0=0.5$ and a varying pumping parameter $\zeta=-0.5$ (solid line),
$\zeta=-0.3$ (dashed line): (a) $w(t)$; (b) $z(t)$.

\vspace{5mm}

{\bf Fig.8.} Transformation of an oscillatory regime of motion for $w(t)$ to
a pulsing one, under $\gm=0.01$, when changing initial conditions: 
(a) $w_0=0.001,\; z_0=-0.1$; (b) $w_0=0.01,\; z_0=-0.1$.

\vspace{5mm}

{\bf Fig.9.} Pulsing regime of motion for $\gm=0.1$, with initial conditions
$w_0=0.01$ and $z_0=-0.5$: (a) $w(t)$; (b) $z(t)$.

\vspace{5mm}

{\bf Fig.10.} Pulsing regime of motion for $\gm=0.01$, with initial conditions
$w_0=0.01$ and $z_0=0.5$: (a) $w(t)$; (b) $z(t)$.

\vspace{5mm}

{\bf Fig.11.} Change in the behaviour of the function $w(t)$, with the same
initial conditions $w_0=0.001,\; z_0=-0.5$, under the variation of the
effective longitudinal relaxation: (a) $\gm=0.1$; (b) $\gm=0.001$.

\vspace{5mm}

{\bf Fig.12.} Dynamics of the function $w(t)$ in the pulsing regime with
different parameters: (a) $\gm=0.1,\; w_0=10^{-6},\; z_0=-0.5$; (b)
$\gm=0.01,\; w_0=0.1,\; z_0=-0.1$.

\vspace{5mm}

{\bf Fig.13.} The coherence coefficient $C_{coh}$, intensity of radiation $I$
in arbitrary units, and polarization $p_z=-z$ versus time for $g=0,\; 
\om_0=200\gm_2$, and $\om=200\gm_2$. The influence of different intial
conditions is analyzed: $z_0=0.475$ (solid line); $z_0=-0.375$ (dashed line);
$z_0=-0.475$ (solid line with crosses).

\vspace{5mm}

{\bf Fig.14.} The same functions as in Fig. 13 for a spin system with 
switched off dipole interactions in the case of $g=0,\; \om_0=200\gm_2$, 
and $z_0=-0.475$. The external alternating field is not in an exact 
resonance, with the frequencies $\om=100\gm_2$ (solid line) and 
$\om=195\gm_2$ (dashed line).

\vspace{5mm}

{\bf Fig.15.} The coherence coefficient $C_{coh}$ and radiation intensity 
$I$ versus time for $g=0,\; \om_0=200\gm_2,\;\om=205\gm_2$, and $z_0=-0.475$.
Different number of spins is considered: $N=343$ (solid line), $N=125$ 
(dashed line), and $N=27$ (solid line with squares).


\newpage

\begin{thebibliography}{99}
\bibitem{1}
Bloembergen, N. and Pound, R. V., 1954, {\it Phys. Rev.}, {\bf 95}, 8.
\bibitem{2}
Dicke, R. H., 1954, {\it Phys. Rev.}, {\bf 93}, 99.
\bibitem{3}
Zheleznyakov, V. V., Kocharovsky, V. V., and Kocharovsky, Vl. V., 1989,
{\it Phys. Usp.}, {\bf 32}, 835.
\bibitem{4}
Zinoviev, P. V., Samartsev, V. V., and Silaeva, N. B., 1991, 
{\it Laser Phys.}, {\bf 1}, 1.
\bibitem{5}
Andreev, A. V., Emelyanov, V. I., and Ilinski, Y. A., 1993, {\it Cooperative
Effects in Optics} (Bristol: Inst. of Physics).
\bibitem{6}
Benedict, M. G., Ermolaev, A. M., Malyshev, V. A., Sokolov, I. V., and
Trifonov, E. D., 1996, {\it Superradiance--Multiatomic Coherent Emission}
(Bristol: Inst. of Physics).
\bibitem{7}
Combrisson, J., Honig, A., Townes, C., 1956, {\it Compt. Rend.}, 
{\bf 242}, 2451.
\bibitem{8}
Feher, G., Gordon, J., Buehler, E., Gere, E., and Thurmond, C., 1958, 
{\bf 109}, 221.
\bibitem{9}
Chester, P. F., Wagner, P. E., and Castle, J. G., 1958, {\it Phys. Rev.},
{\bf 110}, 281.
\bibitem{10}
Abragam, A., Combrisson, J., and Solomon, I., 1957, {\it Compt. Rend.}, 
{\bf 245}, 157.
\bibitem{11}
Wolfe, J. P. and King, A. R., 1976, {\it Chem. Phys. Lett.}, {\bf 40}, 451.
\bibitem{12}
B\"osiger, R., Brun, E., and Meier, D., 1978, {\it Phys. Rev. A}, {\bf 18},
671.
\bibitem{13}
Kiselev, J. F., Prudkoglyad, A. F., Shumovsky, A. S., and Yukalov, V. I.,
1988, {\it Mod. Phys. Lett. B}, {\bf 1}, 409.
\bibitem{14}
Kiselev, Y. F., Prudkoglyad, A. F., Shumovsky, A. S., and Yukalov, V. I., 
1988, {\it J. Exp. Theor. Phys.}, {\bf 67}, 413.
\bibitem{15}
Bazhanov, N. A., Bulyanitsa, D. S., Zaitsev, A. I., Kovalev, A. I., Malyshev,
V. A., and Trifonov, E. D., 1990, {\it J. Exp. Theor. Phys.}, {\bf 70}, 1128.
\bibitem{16}
Reichertz, L., Dutz, H., Goertz, S., Kr\"amer, D., Meyer, W., Reicherz, G.,
Thiel, W., and Thomas, A., 1994, {\it Nucl. Instrum. Methods Phys. Res. A},
{\bf 340}, 278.
\bibitem{17}
Belozerova, T. S., Henner, V. K., and Yukalov, V. I., 1992, 
{\it Phys. Rev. B}, {\bf 46}, 682.
\bibitem{18}
Belozerova, T. S., Henner, V. K., and Yukalov, V. I., 1992, {\it Comput. 
Phys. Commun.}, {\bf 73}, 151.
\bibitem{19}
Slichter, C. P., 1980, {\it Principles of Magnetic Resonance} (Berlin:
Springer).
\bibitem{20}
Yukalov, V. I., 1995, {\it Phys. Rev. Lett.}, {\bf 75}, 3000.
\bibitem{21}
Yukalov, V. I., 1996, {\it Phys. Rev. B}, {\bf 53}, 9232.
\bibitem{22}
Yukalov, V. I., 1997, {\it Proc. Int. Soc. Otp. Eng.}, {\bf 3239}, 118.
\bibitem{23}
Loiko, N. A. amd Samson, A. M., 1993, {\it J. Exp. Theor. Phys.}, {\bf 104},
2314.
\bibitem{24}
B\"osiger, P., Brun, E., and Meier, D., 1979, {\it Phys. Rev. A}, {\bf 20},
1073.
\bibitem{25}
Brun, E., Derighetti, B., Holzner, R., and Meier, D., 1983, {\it Helv. Phys.
Acta}, {\bf 56}, 825.
\bibitem{26}
Alekseev, K. N., Berman, G. P., Tsifrinovich, V. I., and Frishman, A. M., 
1992, {\it Phys. Usp.}, {\bf 35}, 572.
\bibitem{27}
Seigman, A. E., 1964, {\it Microwave Solid--State Masers} (New York: 
McGraw--Hill).
\bibitem{28}
Yukalov, V. I., 1995, {\it Laser Phys.}, {\bf 5}, 526.
\bibitem{29}
Yukalov, V. I., 1995, {\it Laser Phys.}, {\bf 5}, 970.
\bibitem{30}
Yukalov, V. I., 1997, {\it Laser Phys.}, {\bf 7}, 58.
\bibitem{31}
Yukalov, V. I., 1996, {\it Nucl. Instrum. Methods Phys. Res. A}, 
{\bf 370}, 345.
\bibitem{32}
Okunev, I. S. amd Bazhanov, N. A., 1997, {\it St. Petersburg Inst. Nucl. 
Phys. Preprint}, N2185, Gatchina.
\bibitem{33}
Fokina, N. P., Khutsishvili, K. O., Chkhaidze, S. G., and Lomidze, A. M.,
1995, {\it Fiz. Tverd. Tela}, {\bf 37}, 1910.
\bibitem{34}
Haken, H., 1983, {\it Advanced Synergetics} (Berlin: Springer).
\bibitem{35}
Yukalov, V. I., 1997, {\it Phys. Rev. A}, {\bf 56}, 5004.
\bibitem{36}
Weiss, C. O. and Vilaseca, R., 1991, {\it Dynamics of Lasers} 
(Weinheim: VCH).
\bibitem{37}
Belyanin, A. A., Kocharovsky, V. V., and Kocharovsky, Vl. V., 1997, {\it
Quantum Semiclass. Opt.}, {\bf 9}, 1.
\bibitem{38}
Rorschach, H. E. and Low, F. J., 1960, in {\it Quantum Electronics},
Townes, C. H., ed. (New York: Columbia Univ.), p.177.
\bibitem{39}
Robinson, C., 1995, {\it Dynamical Systems} (Boca Raton: CRC).
\bibitem{40}                                                   
Casati, G. and Chirikov, B., 1995, {\it Physica D}, {\bf 86}, 220.
\end{thebibliography}
\end{document}